\documentclass[aps,pra,twocolumn,superscriptaddress,showpacs,floatfix]{revtex4-1}
\usepackage{graphicx,amsmath,amssymb,amsfonts,xspace,epsfig,float,multirow,color,epstopdf}
\usepackage{dsfont}
\usepackage{caption}
\usepackage{subcaption}
\usepackage[T1]{fontenc}
\usepackage[utf8]{inputenc}
\usepackage{bm}
\usepackage{bbm}
\usepackage[english]{babel}
\usepackage[usenames]{xcolor}
\usepackage{float}
\usepackage{hyperref}
 
\newcommand*\dd{\mathop{}\!\mathrm{d}}

\begin{document} 
 
\title{Dynamical depinning of a Tonks Girardeau gas} 
 
\author{Florian Cartarius} 
\affiliation{Theoretische Physik, Universität des Saarlandes, 66123 Saarbrücken, Germany}
\affiliation{Universit\'e Grenoble Alpes, LPMMC, F-38000 Grenoble, France}
\affiliation{CNRS, LPMMC, F-38000 Grenoble, France} 
\author{Eiji Kawasaki} 
\affiliation{Universit\'e Grenoble Alpes, LPMMC, F-38000 Grenoble, France}
\affiliation{CNRS, LPMMC, F-38000 Grenoble, France}
\author{Anna Minguzzi} 
\affiliation{Universit\'e Grenoble Alpes, LPMMC, F-38000 Grenoble, France}
\affiliation{CNRS, LPMMC, F-38000 Grenoble, France}

\date{\today} 
 
\begin{abstract} 
We study the dynamical depinning following a sudden turn off of an optical lattice for a gas of impenetrable bosons in a tight atomic waveguide. We use a Bose-Fermi mapping to infer the exact quantum dynamical evolution. At long times, in the thermodynamic limit, we observe the approach to a non-equilibrium steady state, characterized by the absence of quasi-long-range order and a reduced visibility in the momentum distribution. Similar features are found in a finite-size system  at times corresponding to half the revival time, where we find that the system approaches  a  quasi-steady state  with a power-law behaviour. 
\end{abstract}  
 
 
\maketitle 
 
\section{Introduction}  
The realization of isolated quantum systems has become an experimental reality with ultracold quantum gases. In the experiments, the atoms are confined by magnetic or optical means in ultra-high vacuum conditions and are cooled down to quantum degeneracy in  virtually isolated conditions for a sufficiently long time with respect to the typical time scales for their dynamics. The lifetime of ultracold gases is limited to few seconds, the main decay mechanisms being inelastic three-body and spin-changing collisions.

The study of the dynamics following a quantum quench in isolated quantum systems allows to address fundamental questions in quantum many-body systems (see eg~\cite{Calabrese06}  as well as \cite{Sotiriadis2014} and references therein). 
In one-dimensional integrable systems a relevant issue is the absence of thermalization as oserved in the paradigmatic quantum Newton's cradle experiment \cite{Kinoshita06}. The concept of Generalized Gibbs Ensemble has been introduced and developed to describe the state  of integrable systems at long-times \cite{Rigol07,Cazalilla06,Calabrese07,Cramer08,Barthel08,Iucci09,Calabrese11,Cazalilla12,Mossel12,Caux12,collura13}. 
 
Optical lattices allow to enhance the effects of interactions and explore strongly correlated phases \cite{rmp_dalibard}. Bosons confined in optical lattices display a rich phase diagram. For deep lattices, ie  with lattice depth considerably larger than the recoil energy, the bosons are described by the Bose-Hubbard model and display a superfluid to Mott-insulator transition at increasing the ratio of on-site interaction to tunnel energy \cite{fisher,zoller,bloch}.  Numerical and analytical approaches have also been employed to study interaction quenches in the Bose-Hubbard model (see eg \cite{Kollath07,Dziarmaga14}). In one dimension, for an arbitrarily weak lattice potential commensurate with the particle number density, the bosons still display  a  gapped, insulating  phase, provided that  interactions are sufficiently strong. This  pinning transition is well accounted for within the Luttinger-liquid description and a renormalization group analysis of the effect of the lattice, which yields a sine-Gordon Hamiltonian \cite{zwerger}.  Various types of quantum quenches in the sine-Gordon model have been studied with different techniques eg. refermionization \cite{Iucci10}, flow equations \cite{Sabio10}, and form factors \cite{Fioretto10}.

Ultracold atomic gases confined by tight waveguides have been experimentally realized and the strongly interacting regime has been reached and characterized in detail \cite{Paredes04,Kinoshita04,VanDruten08,Palzer09}. In this one-dimensional geometry, the pinning transition has  been experimentally observed for atoms subjected to a longitudinal weak optical lattice \cite{Haller10}.

In this work, we consider specifically  a  quench across the pinning transition. In detail, we consider as initial state  a one-dimensional bosonic gas in the limit of infinitely strong interactions (or Tonks-Girardeau gas), confined by an optical lattice and with a number density commensurate with the lattice spacing \cite{Buljan}. We follow its time evolution following a sudden turn-off of the lattice while keeping the one-dimensional waveguide still present. This allows us to study the quench from  an initially pinned, insulating ground state to  an out-of-equilibrium depinned state.  We describe the dynamical evolution of the system at arbitrary times using an exact mapping solution due to Girardeau \cite{Gir1960,girardeau_timedep}. This exact solution allows us to obtain the full dynamical solution for the quantum dynamics, going beyond the low-energy Luttinger-liquid model or conformal field theory approaches. Our method allows us also to study the approach in time to the non-equilibrium steady state. Finally, we focus  on the experimentally relavant condition of a finite-size system, choosing a  geometry that is amenable to experimental realization with ultracold atoms, eg by implementing a one-dimensional box-potential confinement \cite{Gaunt12} to which an optical lattice is superimposed.

\section{Model and exact solution} 

\begin{figure*}[th]
	\centering 
	\includegraphics[width=0.7\textwidth]{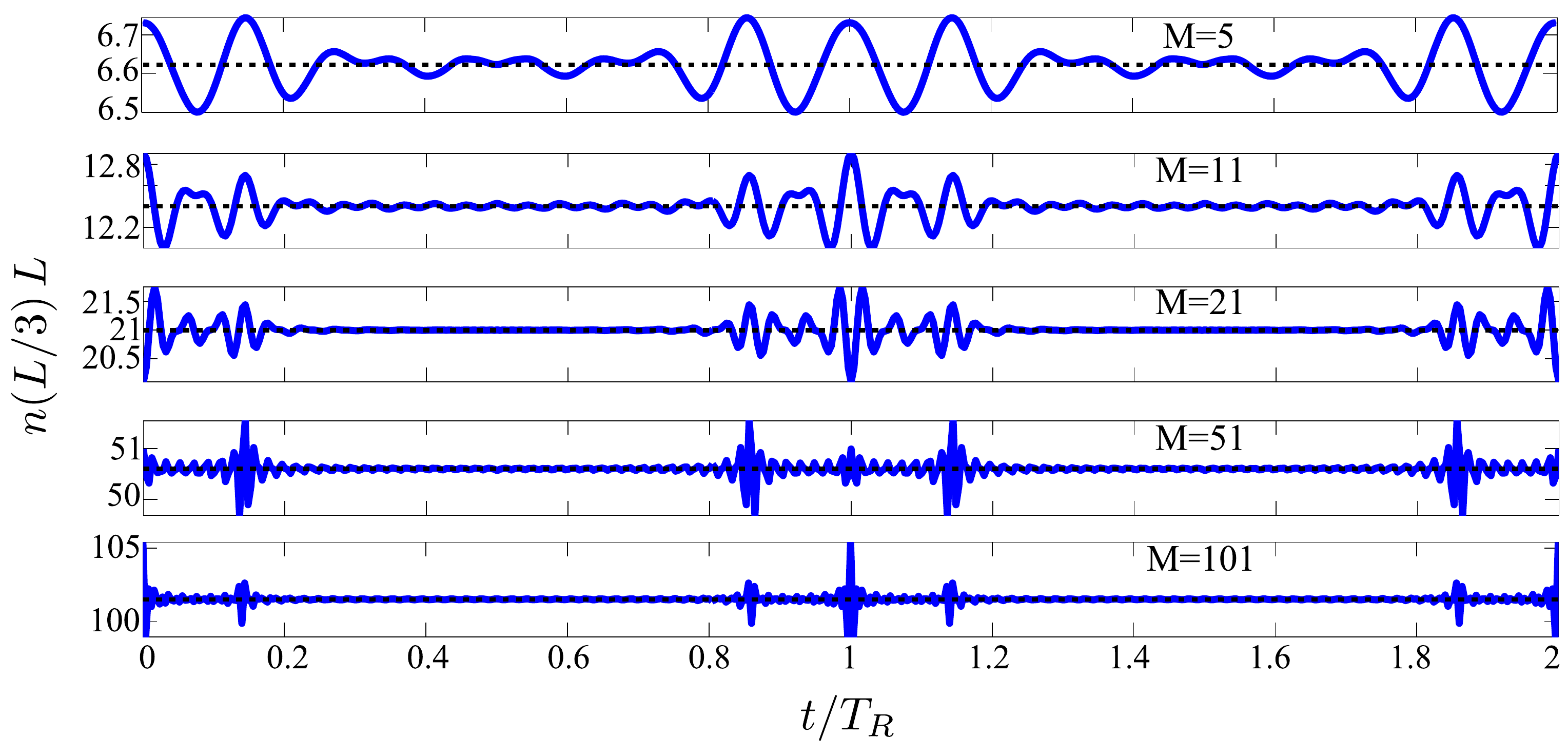} 
	\caption{\label{Fig:density}  
		(Color online) Time evolution of the particle density $n(L/7,t)$ at position $x=L/7$ as a function of time $t$ for a commensurately filled lattice for various values of boson numbers as indicated in each panel. The horizontal dashed lines correspond to the (quasi) steady-state prediction in Eq.~(\ref{n_SS}). 
	} 
\end{figure*} 

We consider  $M$  bosons of mass $m$ at zero temperature, confined by a longitudinal box trap of size $L$. They are described by the Hamiltonian  
\begin{equation} 
{\cal H}=\sum_j\left[ -\frac{\hbar^2}{2m} \frac{\partial^2}{\partial x_j^2} + V(x_j)\right]+ g \sum_{j<\ell} \delta(x_j-x_\ell). 
\label{eq:ham} 
\end{equation} 
The atoms are subjected to an optical lattice and to a box potential, ie  $V(x)=V_L \cos^2(k_L x) + V_b(x)$, where  $k_L=M \pi/L$ and the box trap $V_b(x)$ is described by imposing hard-walls boundary conditions on the interval $[0,L]$. The number of bosons is chosen such as to ensure unitary filling of the lattice.
The interactions among the atoms is modelled  by contact interactions of strength $g$.    In the following we shall focus on the impenetrable-boson or Tonks-Girardeau limit corresponding to the limit $g\rightarrow \infty$. This amounts to replacing the interaction term in the Hamiltonian  (\ref{eq:ham})  by the   cusp condition  $\Psi(...x_j=x_\ell...)=0$, imposing the vanishing of the wavefunction at contact for each pair of  particles $\{j,\ell\}$. 

At time $t=0^-$ we assume that the gas is at equilibrium in the optical lattice. We study the time evolution of the gas following a sudden quench of the lattice amplitude $V_L$ to zero. The exact dynamics is described by the time-dependent Bose-Fermi mapping, stating that the time evolution of the  bosonic wavefunction $\Psi_{B}(x_1...x_M,t)$ can be obtained in terms of the one of a non-interacting Fermi gas subjected to the same time-dependent external potential according to
\begin{equation} 
\label{Eq:psi_TG} 
\Psi_{B}(x_1...x_M,t)=\Pi_{1\le j<\ell\le M} {\rm sign}(x_j-x_\ell) \Psi_{F}(x_1,x_2..,x_M,t), 
\end{equation} 
where $\Psi_{F}(x_1,x_2..,x_M,t) = \frac{1}{\sqrt{M!}} \det[\psi_j(x_k,t)]$. Note that the solution (\ref{Eq:psi_TG}) satisfies the cusp condition at all times. 
The single-particle orbitals $\psi_j(x_k,t)$ are the solution of the time-dependent one-body  Schrödinger equation 
\begin{equation} 
 -\frac{\hbar^2}{2m}\partial_x^2 \psi_j(x,t) + V_b(x) \psi_j(x,t)= i \hbar \partial_t \psi_j(x,t),
\label{Eq:sp} 
\end{equation}  
As initial condition $\psi_j(x,0)$ we take the equilibrium single-particle problem in  the presence of the lattice, corresponding to a  Mathieu equation with hard walls boundary conditions, 
\begin{equation} 
 -\frac{\hbar^2}{2m}\partial_x^2 \psi_j(x) + [V_L \cos^2(k_L x)+V_b(x)] \psi_j(x)= E_j \psi_j(x). 
\label{Eq:sp_eq} 
\end{equation} 
In the following it will be useful to scale all the energies in units of the recoil energy $E_R=\hbar^2 k_L^2/2m$, and set $\lambda=V_L/E_R$. The solution of Eq.(\ref{Eq:sp_eq}) is given by a  generalization of Mathieu functions. It amounts to search for a solution of the form $\psi_j(x)=\sqrt{2/L}\sum_n b_n^{(j)} \sin(n \pi x/L)$ and determine the coefficients $ b_n^{(j)}$. Substitution onto the Schrödinger equation (\ref{Eq:sp_eq}) yields the linear algebra problem 
\begin{equation} 
\sum_{n}  \sin(n k_L x)(b_n^{(j)} (n^2/M^2-a) +q (b_{n+2M}^{(j)}+b_{n-2M}^{(j)}))=0 
\label{diag_mathieu} 
\end{equation} 
where $a=E_j/E_R-\lambda/2$, $q=\lambda/4$. 
Equation (\ref{diag_mathieu}) corresponds to an eigenvalue problem on a semi-infinite matrix. For the ground state of the TG gas we are interested in the first $M$ eigenvalues  and eigenvectors. These are obtained by numerical diagonalization,  performing a truncation to a matrix size $S\gg M$. The time evolution of the single-particle orbitals after the sudden depinning is then readily given by 
\begin{equation} 
\psi_j(x,t)=\sqrt{2/L}\sum_n b_n^{(j)} e^{-i \varepsilon_n t/\hbar}\sin(n \pi x/L)
\label{eq:psij}
\end{equation} 
where $\varepsilon_n=\hbar^2 (n \pi/L)^2/2m$. 
 
\section{Time evolution of the density profiles} 
In order to explore the post-quench dynamics we analyze the time evolution of various observables. We consider first the time evolution of the density profile. This is obtained, using the   
Bose-Fermi mapping, as the one of the corresponding Fermi gas,  
\begin{equation} 
n(x,t)=\sum_j^M |\psi_j(x,t)|^2 .
\end{equation} 
Substitution of the explicit solution for the lattice problem yields 
\begin{eqnarray} 
n(x,t)&=&\frac 2 L \sum_j^M \sum_{n,n'}e^{- i (\varepsilon_n-\varepsilon_{n'})t/\hbar} b_n^{(j)}  b_{n'}^{(j)} \nonumber \\
&&\times \sin(n \pi x/L)\sin(n' \pi x/L). 
\label{time-density} 
\end{eqnarray} 
The time evolution of the density is shown in Fig.~\ref{Fig:density} for various numbers of bosons and various fixed sizes, at constant filling of one boson per site.  Recurrences are clearly visible, as expected for a finite-size system. While one could estimate as trivial recurrence time   $T=2 \pi \hbar/\varepsilon_1$, we note that the density profiles show revivals at earlier time $T_R=T/4M$. This property is specific to our choice of system and initial state. The time evolution  in Eq.~(\ref{time-density}) is determined by the energy difference $\varepsilon_n-\varepsilon_{n'}\propto (n^2-n'^2)$; the most important contribution in the coefficients $b_n^{(j)}$ is for $n'-n=0$ mod $2M$, as one can infer from the numerical solution  of Eq.~(\ref{diag_mathieu}) as well as from a perturbative approach \cite{Buljan} at weak lattice strength, thereby yielding the observed recurrence time $T_R$.

\section{The non-equilibrium steady state} 
\begin{figure} 
\centering 
\includegraphics[width=1\linewidth]{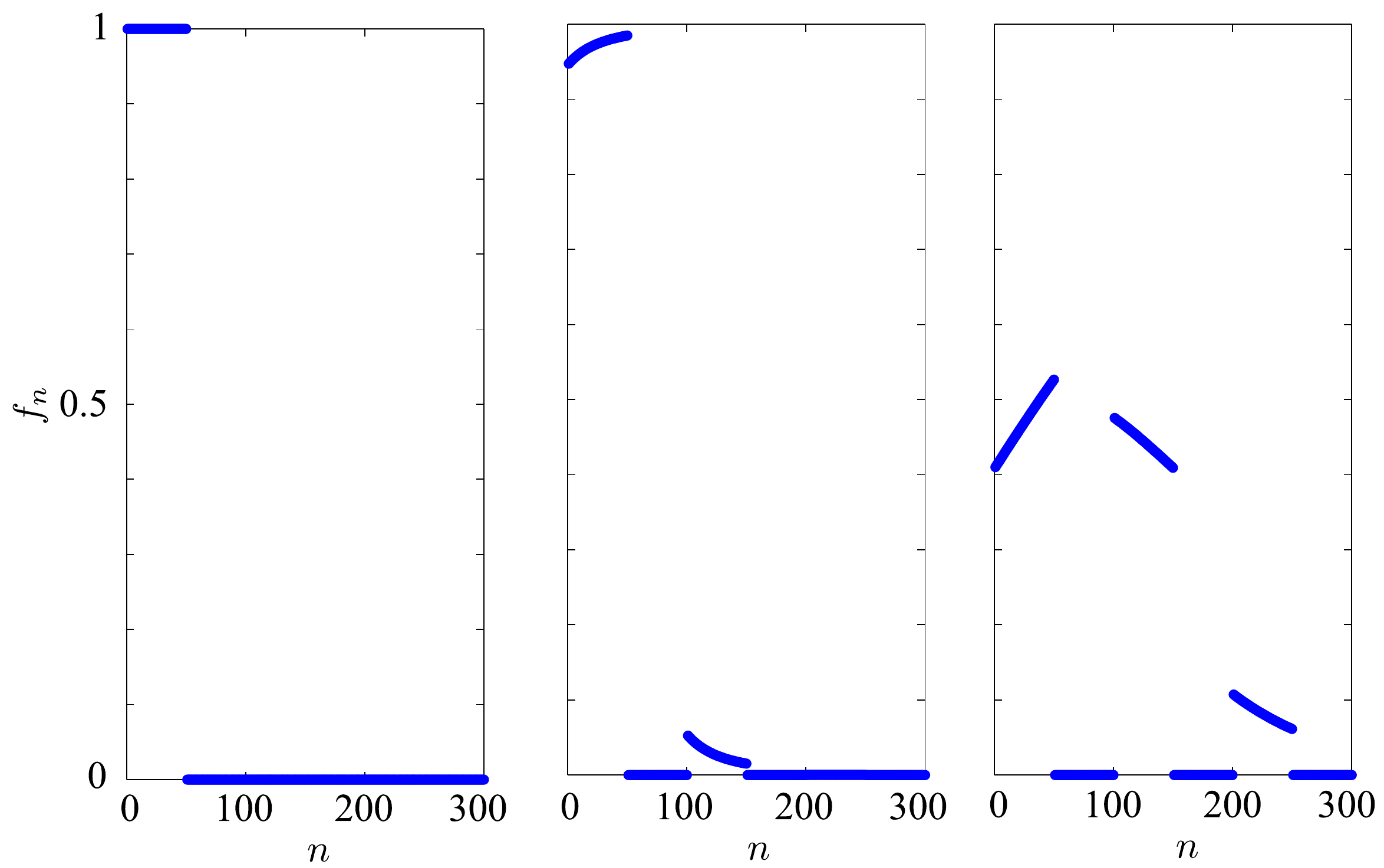}
\caption{\label{Fig:occupations}  
(Color online) Occupation numbers $f_n$ as a function of the quantum number $n$ for $M=50$ particles, for various values of the dimensionless lattice strength, from left to right, $\lambda=0,4, 50$.
} 
\end{figure}

 For a value of  $M$ sufficiently large to be outside the mesoscopic regime of a few particles ($M\ge 11$ in our case) and for sufficiently long times (ie at about half revival time) we observe in Fig.~\ref{Fig:density} that the system tends to a (quasi) steady state -- as we shall denote this state for a finite-size  system.~\footnote{According to our numerical solution, the density profile is very close to the steady-state prediction~(\ref{n_SS}) except for travelling wiggles due to reflections against the walls.} This state tends to a truly steady state once the thermodynamic limit is taken (see Section \ref{sec:nk} below for details). The (quasi) steady state  can be well described by neglecting the oscillating terms in Eq.~(\ref{time-density}) 
\begin{equation} 
n(x,t) \rightarrow n^{SS}(x)= \frac 2 L \sum_{n=1}^\infty f_n \sin^2(n \pi x/L) 
\label{n_SS} 
\end{equation} 
with nonthermal occupation numbers $f_n$ given by $f_n=\sum_{j=1}^N |b_n^{(j)}|^2$.
The above result (\ref{n_SS}) is illustrated as horizontal dashed lines in  Fig.\ref{Fig:density}.  In the absence of the lattice $f_n$ is given by a Fermi distribution at zero temperature. At increasing height of the  initial lattice  the (quasi) steady state is characterized by the occupation of more and more excited bands, as shown in Fig.~\ref{Fig:occupations}.  {\color{black} We notice that the occupation numbers $f_n$  vanishes for $n=M+1$ to $2M$, and similarly for higher excited levels corresponding to even bands. This is due to the fact that the optical lattice acts as a backscattering potential creating excitations with wavevector $2k_L$, which corresponds to  $2k_F = 2 M\pi/L$ in our choice of lattice filling.  In energy space, this allows to excite only levels with quantum number difference $\Delta n= 2 M$.  Mathematically, the result follows from the linear algebra problem in equation (5) : the optical lattice gives rise to  off-diagonal terms into the matrix that are a distance 2M from the diagonal. The corresponding eigenvectors have mostly zero components except $b_n^{(j)} \neq 0, \forall n = 2M+1$. Combining this property with the definition of the occupation numbers $f_n$ we obtain the result shown in Fig.2.}

\begin{figure} 
\centering 
\includegraphics[width=1\linewidth]{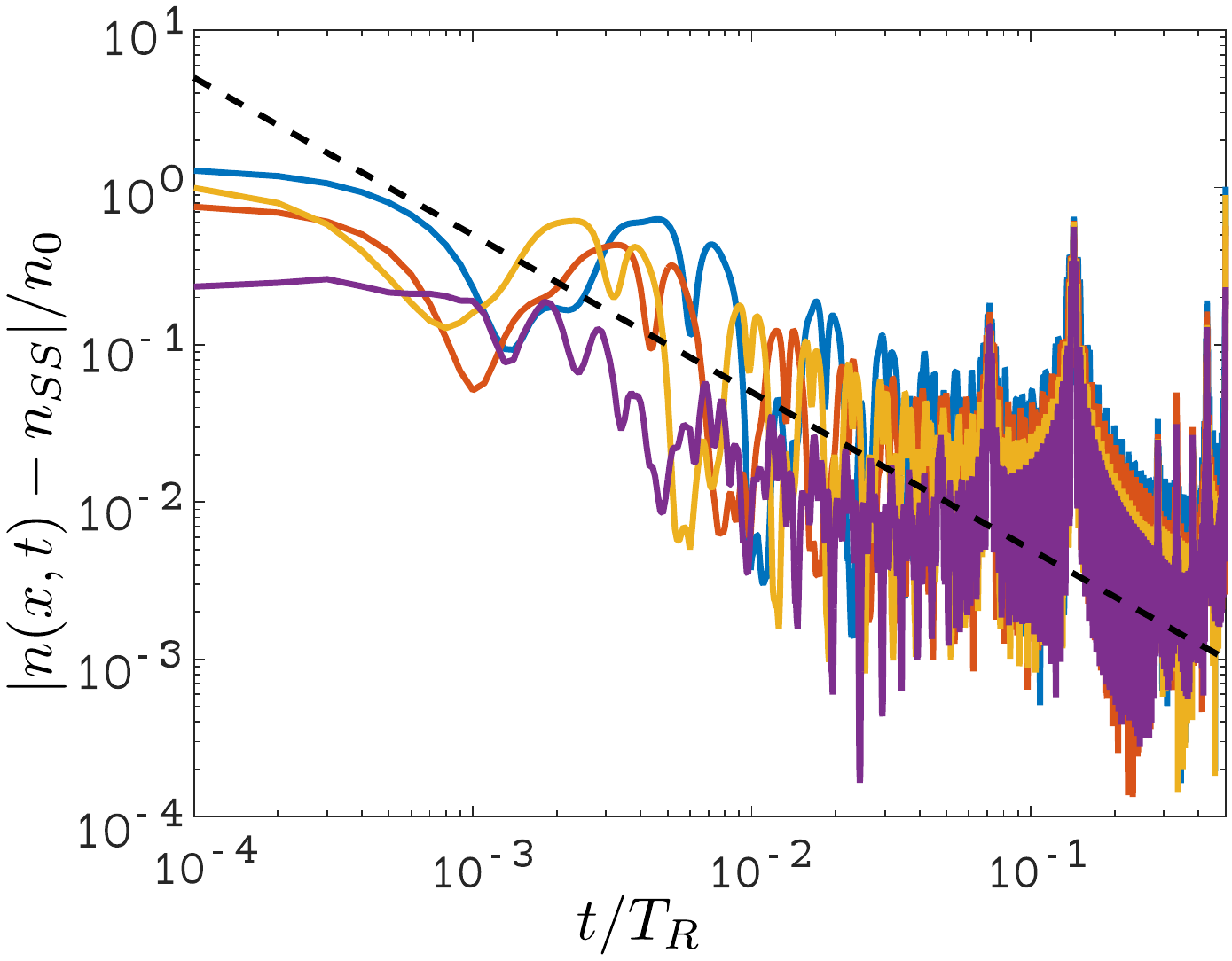}
\caption{(Color online) Approach to nonequilibrium steady-state: time evolution (time in units of $T_R$) in double logarithmic scale for the difference between the particle density and its corresponding steady-state value   $|n(x,t)-n^{SS}(x)|$ (in units of $n_0 = M/L$), evaluated at  $x= L/7$ and for $\lambda=40$, for 81 (blue), 111 (red), 151 (orange) and 201 particles (purple). The particle density has been time averaged over a short time interval $\Delta t=0.001\, T_R$  in order to decrease the noise in the figure. The dashed line indicates the $1/t$ power law decay. \label{Fig:approach}  
} 
\end{figure}

Furthermore, our exact solution allows also to explore the approach to steady state. By an extensive analysis of various systems sizes, we have found a power-law approach to the steady state ie, $n(x,t)-n^{SS}(x)\sim 1/t$ as times approaches $T_R/2$, which corresponds to the large-time limit in our finite-size system. This is illustrated in Fig.\ref{Fig:approach} for various values of particle number $M$.
 
\section{First-order coherence and momentum distribution} 
\label{sec:nk} 
In order to further characterize the properties of the (quasi)  steady state we study the  time-dependent one-body density matrix $\rho_1(x,y,t)=M \int \dd x_2..\dd x_N \Psi_B^*(x,x_2,...x_M,t) \Psi_B(y,x_2,...x_M,t)$. This allows to determine the coherence properties of this state and in particular the presence of quasi-off-diagonal long-range order (QODLRO). Furthermore, this allows to obtain the momentum distribution of the gas $n(k,t)=\int \dd x \int \dd y \rho_1(x,y,t)e^{i k (x-y)} $, which is experimentally accessible with a high precision (see eg \cite{fang2012}). 
 
Following the approach of Ref.~\cite{PezBul07} the one-body density matrix of a time-evolving Tonks-Girardeau gas is given by 
\begin{equation} 
\rho_1(x,y,t) = \sum_{j,l = 1}^M \psi_j^*(x,t) A_{jl}(x,y,t) \psi_l(y,t). 
\label{Eq:pezer}
\end{equation} 
where the matrix $A(x,y,t)= ( P^{-1})^T \text{det} P$, $ P(x,y,t) = {\mathbbm 1} -  Q$, with  $Q_{jl}(x,y)  = 2 \, {\rm sign} (y-x) \; \int_x^y \dd x' \psi_j^* (x',t) \psi_l (x',t)$. For our specific case,  the matrix elements $ Q_{jl} (x,y)$ are readily evaluated analytically using Eq.~(\ref{eq:psij}). 
 
\begin{figure*}[th] 
\centering 
\includegraphics[width=1\textwidth]{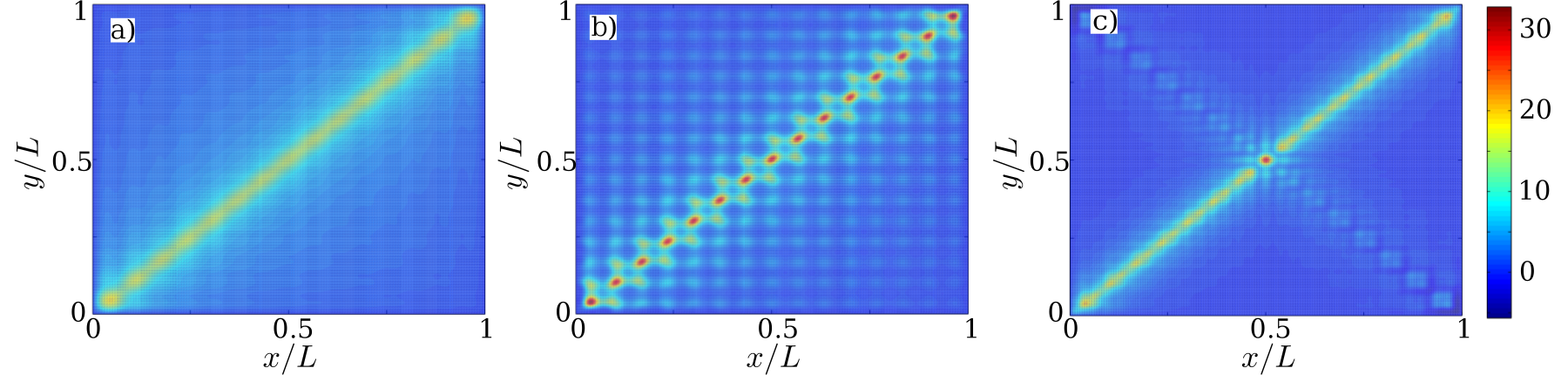}
\includegraphics[width=1\textwidth]{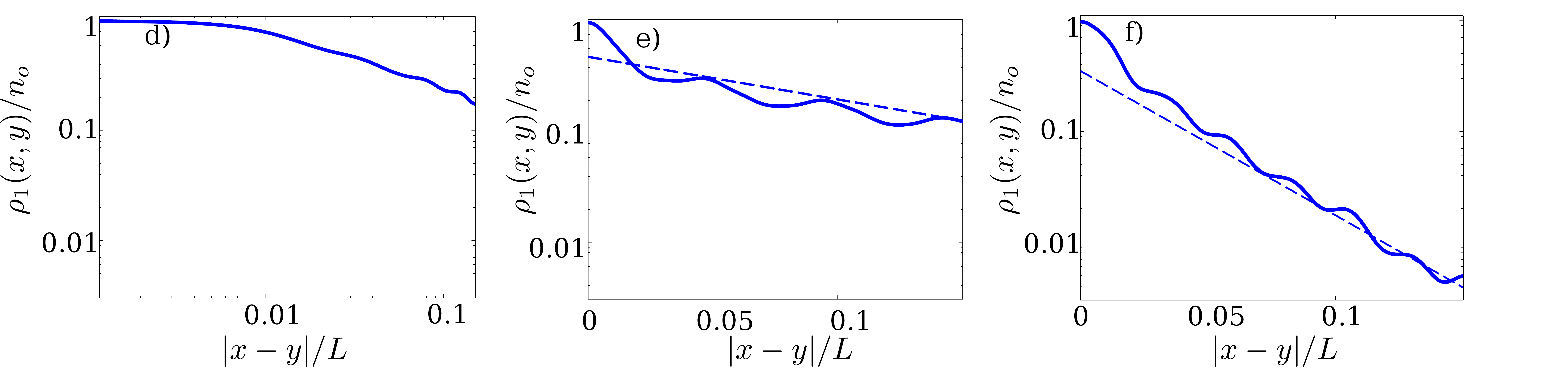}
\caption{\label{Fig:densitymatrix}  
(Color online) Real part of the one-body density matrix $\rho_1(x,y)$ in units of $n_0=M/L$ as a function of the coordinates of the spatial coordinates $x$ and $y$ for $M=15$ bosons. Top panels: top view, bottom panels: corresponding cuts at fixed $x+y=L/2$ as a function of the relative distance $|x-y|$. a) and d) equilibrium state in absence of the lattice. b) and e) equilibrium state in presence of the lattice, with lattice strength $\lambda = 10$. c) and f) depinned quasi steady state at time $t=T_R/2$ after a quench of the lattice to zero. The dashed line in panel e) is a guide to the eye and in panel f) indicates the exponential decay $e^{- 2 n |x-y|}$ predicted in Eq.(\ref{rho1_ss}).
} 
\end{figure*}

The resulting density matrices are illustrated in Fig.\ref{Fig:densitymatrix}, { for 15 bosons.} As compared with the equilibrium case in absence of the lattice (panels a) and d)), the effect of the lattice at time $t=0$ is a pinning  along the diagonal $x=y$ and and a reduction of the off-diagonal coherences,  as evident from the sections taken along the direction $x=-y$ (panels e) and f)). As a main result, we find that  the (quasi) steady state at time $t=T_R/2$ displays no QODLRO, and the one-body density matrix decays exponentially \footnote{ This result was checked by studying the density matrices for systems up to 31 bosons and an initial lattice strength of $\lambda = 10$. We have also checked that a similar behaviour is found for times close but not equal to $T_R/2$.}.
This is especially striking since the system is evolving in a homogeneous box, ie it is depinned: While the corresponding equilibrium state in the box displays  QODLRO, ie the well-known power-law decay with power exponent -1/2, as shown in panel d) of Fig.~\ref{Fig:densitymatrix}, quite remarkably, the non-equilibrium aspect of the gas influences dramatically its coherence properties. This feature is found quite generally in integrable models, and has been first predicted by conformal-field methods for the state of the system at long times following a quench across a quantum critical point \cite{Calabrese06}. 

For weak lattice strength, the exponential decay of the one-body density matrix for the (quasi) steady state can be analytically obtained: the main contribution to the weights $b_n^{j}$ is given by the term $\delta_{n,j}$, yielding for the matrix elements $P_{i,j}=\delta_{i,j} [ 1-2 |(x-y)/L - [\sin(2 \pi j x/L)-\sin(2 \pi j y/L)]/(2 \pi j)| ]$. At large relative distances one may set $P_{i,j}\simeq \delta_{i,j}(1-2 |x-y|/L)$, and thereby obtaining $A=  {\mathbbm 1} (1- 2|x-y|/L)^{(N-1)}$.
Taking the thermodynamic limit $M\rightarrow \infty$ and $L\rightarrow \infty$  at fixed $n=M/L$  we find $A= {\mathbbm 1} \exp(-2 n |x-y|)$ and finally using (\ref{Eq:pezer}) we obtain the bosonic quasi-steady state one-body density matrix
\begin{equation}
\rho_1(x,y,t\rightarrow \infty) =\rho_{1F}(x,y,t\rightarrow \infty) \; e^{-2 n |x-y|},
\label{rho1_ss}
\end{equation}
where $\rho_{1F}(x,y,t)= \sum_{j= 1}^M \psi_j^*(x,t) \psi_j(y,t)$ is the fermionic one-body density matrix. This result coincides with the one obtained in  \cite{collura13} for the release of a harmonically trapped TG gas onto a ring: as expected, boundary conditions do not affect the result in the thermodynamic limit. Notice however the different way the steady state is reached: while in Ref.~\cite{collura13} the long-time state is obtained by summing up all the periodically repeated images on the ring, in our case it is obtained by multiple reflections at the boundaries. Notice that our result can then be linked to the concept of  Generalized Gibbs ensemble (GGE) for the thermodynamic limit of our model:  as in \cite{collura13}, the average of the fermionic occupation numbers $\langle c^\dagger_k c_k \rangle$, obtained from the initial fermionic one-body density matrix according to $\langle c^\dagger_k c_k \rangle=\int dx \int dy \rho_{1F}(x,y,0) e^{i k (x-y)} /(2\, \pi) $ are conserved in the time evolution and can be used to determine the Lagrange multipliers $\lambda_k$ defining the density matrix of the system at long times: $\hat \rho_{GGE}\sim \exp(-\sum_k \lambda_k c^\dagger_k c_k )$. 

\begin{figure}
\centering 
\includegraphics[width=1\linewidth]{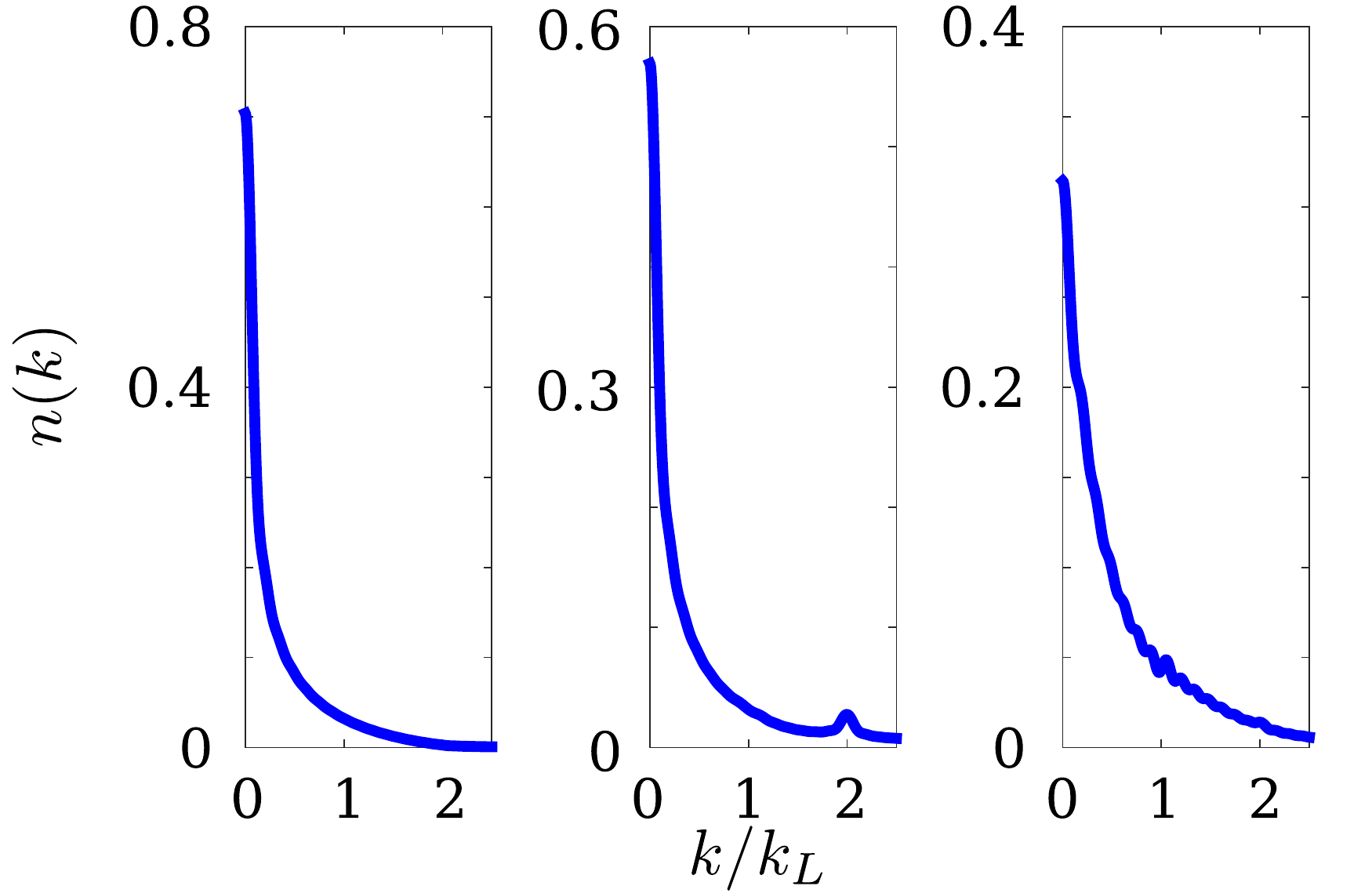}  
\caption{\label{Fig:nk}  
(Color online) Momentum distribution $n(k)$ as a function of wave vector $k$ for $M=15$ particles. Left: equilibrium state in absence of the lattice, center: equilibrium state in presence of a lattice with $\lambda =10$, right: quasi steady state at time $t=T_R/2$ after a quench of the lattice to zero. 
} 
\end{figure}

From the knowledge of the bosonic one-body density matrix $\rho_1(x,y,t)$ we finally obtain the exact  momentum distribution of the gas. We stress that our exact approach allows us to cover all the ranges of momenta, beyond the low-momentum region accessible by conformal-field methods.   As shown in Fig.\ref{Fig:nk}, the momentum distribution of the (quasi) steady state is considerably different from the equilibrium ones both in absence and in presence of the lattice:  it displays a considerably reduced intensity at low momenta, and  does not show the typical backscattering peak at $k=2 k_F$ found for the equilibrium gas in the presence of the lattice.

\section{Conclusions and outlook}
In conclusion, we have studied the exact time evolution of a Tonks-Girardeau gas following its sudden depinning off a weak optical lattice. We have identified a suitable long-time limit where a non-equilibrium steady state is reached in the thermodynamic limit, and we have shown a power-law approach to the steady state. Furthermore, we have shown that this state is characterized by the absence of quasi-long-range order ie  an exponential decay of one-body correlations, in agreement with the predictions of the Generalized Gibbs Ensemble. Our numerical  analysis for a system of finite size shows that this scenario could be reached with experimentally realistic numbers of bosons in a  tight atomic waveguide, and that the time-dependent momentum distribution yields relevant information about this state. Our work opens to the study of the details of the quench depinning dynamics of a Lieb-Liniger gas at arbitrary interactions and to further tests of the GGE hypothesis with hard walls boundary conditions (see eg \cite{Goldstein15}).

\acknowledgments We acknowledge discussions with I.~Bouchoule,  G. Carleo, J.-S. Caux and G. Morigi. AM acknowledges support from the Handy-Q ERC grant no. 25860 and from the ANR project Mathostaq ANR-13-JS01-0005-01. 

\bibliography{cit}
 
\end{document}